\documentclass[conference,a4paper]{IEEEtran}

\newlength{\simulationskip}
\usepackage{xcolor}
\usepackage{balance}
\usepackage{amsmath}
\usepackage{url}

\ifCLASSINFOpdf
  \usepackage[pdftex]{graphicx}
\else
  \usepackage[dvips]{graphicx}
\fi
\ifCLASSOPTIONcompsoc
 \usepackage[caption=false,font=normalsize,labelfont=sf,textfont=sf]{subfig}
\else
 \usepackage[caption=false,font=footnotesize]{subfig}
\fi
\IEEEoverridecommandlockouts

\begin{document}
%
\title{Fast Full-Wave Simulation of Indoor RSS Maps for Pre-Measurement Validation in Device-Free Localization \thanks{Funded by the European Union. Views and opinions expressed are however those of the author(s) only and do not necessarily reflect those of the European Union or European Innovation Council and SMEs Executive Agency (EISMEA). Neither the European Union nor the granting authority can be held responsible for them. Grant Agreement No: 101099491 (project HOLDEN).}}

\author{
	\IEEEauthorblockN{Federica Fieramosca\IEEEauthorrefmark{1}, Anastasia Maiolli\IEEEauthorrefmark{1}, Alexander H. Paulus\IEEEauthorrefmark{2}, Stefano Savazzi\IEEEauthorrefmark{3}, Michele D'Amico\IEEEauthorrefmark{1}}
	
\IEEEauthorblockA{\IEEEauthorrefmark{1} DEIB, Politecnico di Milano, Milan, Italy, \{federica.fieramosca, michele.damico\}@polimi.it}
	
\IEEEauthorblockA{\IEEEauthorrefmark{2}CIT, Technical University of Munich, Munich, Germany, \{a.paulus\}@tum.de
    }

\IEEEauthorblockA{\IEEEauthorrefmark{3} Consiglio Nazionale delle Ricerche (CNR), IEIIT Institute, Milan, Italy, \{stefano.savazzi\}@cnr.it}
	}



\maketitle

\begin{abstract}
Human localization is gaining momentum in security, healthcare, logistics, and smart spaces applications. While global navigation systems are unreliable indoor, device-free (a.k.a. passive) localization methods that exploit human-induced perturbations of radio propagation can be effectively used. This paper investigates the use of a compact full-wave electromagnetic (EM) setup as a fast and reliable tool to simulate indoor Wi-Fi propagation for human sensing. The goal is to provide a practical baseline for validating simplified propagation models, such as diffraction-based descriptions, and to reduce the need for costly measurement campaigns. Two-dimensional attenuation maps from received signal strength are generated and compared in controlled environments, focusing on attenuation statistics and interference patterns. The simulations reproduce the main spatial features, though discrepancies remain due to simplified material characterization. Diffraction-aware refinements are proposed to mitigate these effects. Overall, the approach provides an efficient pre-measurement reference to support device-free system design and to guide experimental planning.
\end{abstract}

\vskip0.5\simulationskip

%

\section{Introduction}

Indoor localization through device-free (or passive) methods \cite{paper2,paper14, paper27} is a key technology for human behavior analysis in healthcare and industry domains \cite{shit-2019}. For instance, it can be implemented with camera-based~\cite{opti_track}, laser-based~\cite{htcvive_3}, and inside-out optical systems~\cite{Htcultimate}. At lower frequencies and relying on additional devices to be carried, or attached, to a target of interest, radio-frequency identification~\cite{rfidsensing1,rfidsensing2} and ultra-wideband systems~\cite{uwbsensing} can provide less accurate position information regardless of lighting conditions and partially through walls as well. However, device-free localization and counting ~\cite{IoT25_Fie} at GHz frequencies is possible e.g. via radio tomographic imaging~\cite{paper27,paper19}, by exploiting proper modeling of wave propagation phenomena within the environment where a person (i.e., a target) is observed. In WiFi bands, even moderate sized office rooms can represent electrically large scenarios which make efficient EM simulations, e.g., utilizing ray tracing, uniform-theory-of-diffraction (UTD) formulations~\cite{paper26}, as well as accelerated full-wave electromagnetic solvers, recommendable.
%
%
Among these, full-wave analysis is generally the most complete and reliable when geometry, materials, and antennas are specified accurately~\cite{Fieramosca2023FloorEffects,paper33_Fieramosca2024FullWave}. 
However, the fine details of furniture and the vast variety of material properties may render a detailed scene characterization necessary and lead to significant computational time.

To support the development of simplified propagation models (e.g., diffraction-based descriptions~\cite{paper25,paper26}), we investigate a compact FEKO~\cite{Feko} setup as a fast pre-measurement reference that reproduces indoor RSS maps with sufficient spatial fidelity without a full CAD reconstruction.

\begin{figure}[!t]
  \centering
  \includegraphics[width=\columnwidth]{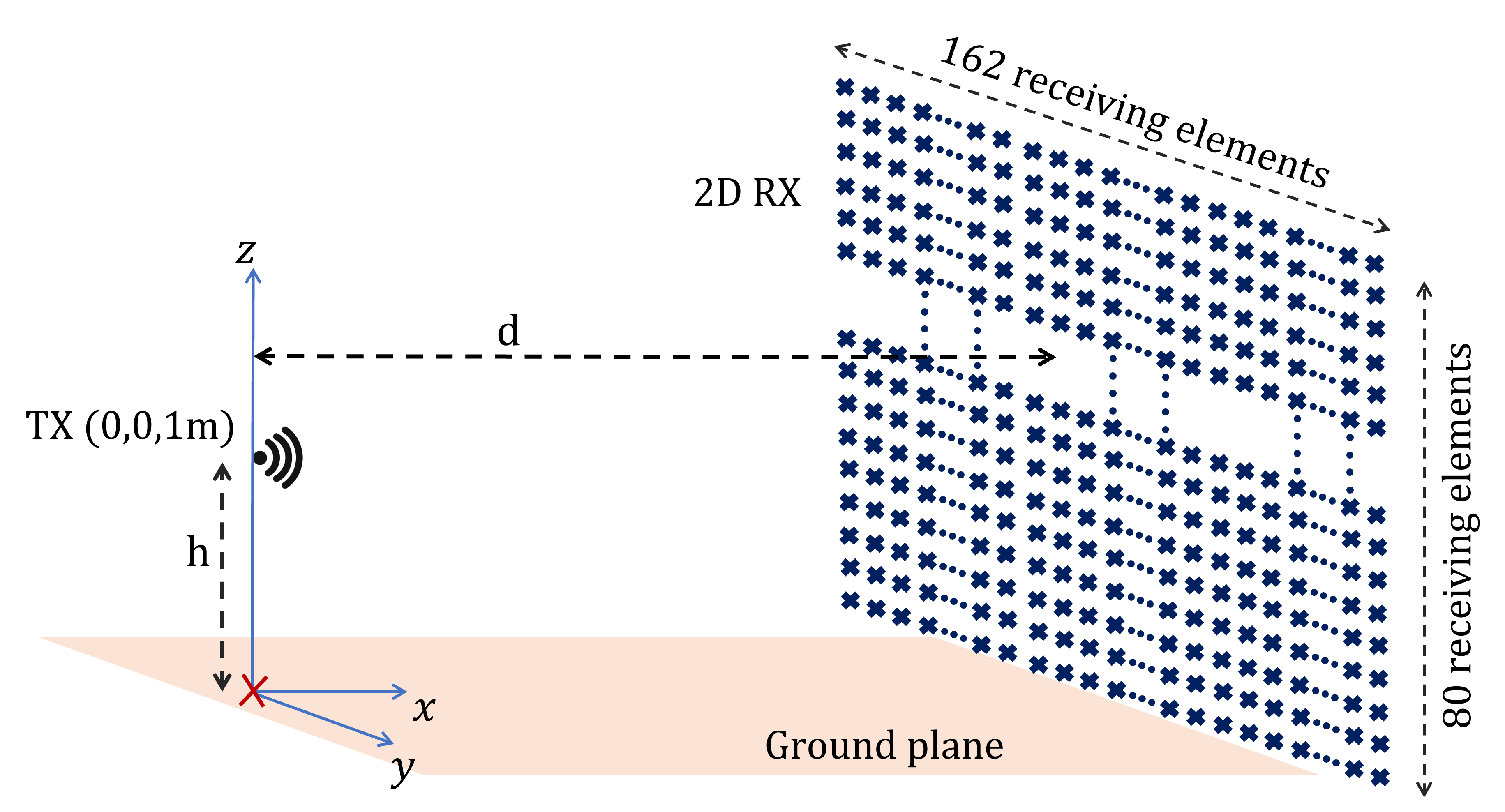}
  \caption{Measurement geometry and receive sampling grid. The Tx is at height $h_{\mathrm{Tx}}{=}1$\,m and $0.10$\,m from the back wall. The Rx is scanned over a vertical plane offset by $0.283$\,m from the side wall; blue marks indicate the $162{\times}80$ sampling points (with an uniform step $\approx3.1$\,cm). Axes and ground plane (concrete) match the modeling in Sec.~\ref{sec:method}.}
  \label{fig:setup-schematic}
\end{figure}

The proposed study is based on an experimental campaign~\cite{paulus_apwc25} that exploits dense, phase-stable Received Signal Strength (RSS) measurements. Compared with~\cite{paulus_apwc25}, a different exploitation of the data is performed. A set of 2D RSS attenuation maps for specific target placements is obtained and used to: 
(i) assess a compact FEKO simulation for synthetic reproduction of measurements, and (ii) enable the injection of image sources via \emph{data-driven} tuning of a single complex wall reflection coefficient by maximizing correlation on the common receiver (Rx) grid. Relative to~\cite{paper33_Fieramosca2024FullWave}, which focused on full-wave analysis of human-body blockage with dense 2D arrays, here we shift the emphasis to room-scale RSS map agreement and lightweight modeling. The proposed processing pipeline (image-source completion and effective reflection tuning) provides an experimental cross-check that supports the pre-deployment design of device-free RF sensing tools. 

According to the geometry highlighted in Fig.~\ref{fig:setup-schematic}, we generate two-dimensional (2D) RSS attenuation maps in FEKO from a minimal scene description and compare them with controlled measurements from~\cite{paulus_apwc25}. Agreement is assessed via interference patterns and map-level metrics (i.e., correlation).  

The paper is organized as follows: Sect.~\ref{sec:method} describes the indoor measurement scenario and methodology, while  Sect.~\ref{sec:free_space} targets the initial full-wave simulations in the free space case. Sect.~\ref{sec:reflections} highlights the proposed data-driven optimization for modeling multipath effects, and Sect.~\ref{sec:results} reports results and discussion. Finally, Sect.~\ref{sec:conc} draws some conclusions.

\section{Measurement Scenario and Simulation}
\label{sec:method}

This section introduces the indoor measurement setup used to build ground-truth reference maps and the corresponding full-wave simulation adopted in FEKO for fast synthetic reproduction.

\subsection{Indoor measurement setup and data}
\begin{figure}[!t]
\centering
\subfloat[]{
        \includegraphics{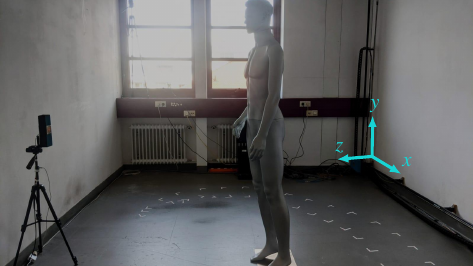} 
		}

\hspace*{1cm}
        \subfloat[]{
            \includegraphics[height=6.75cm]{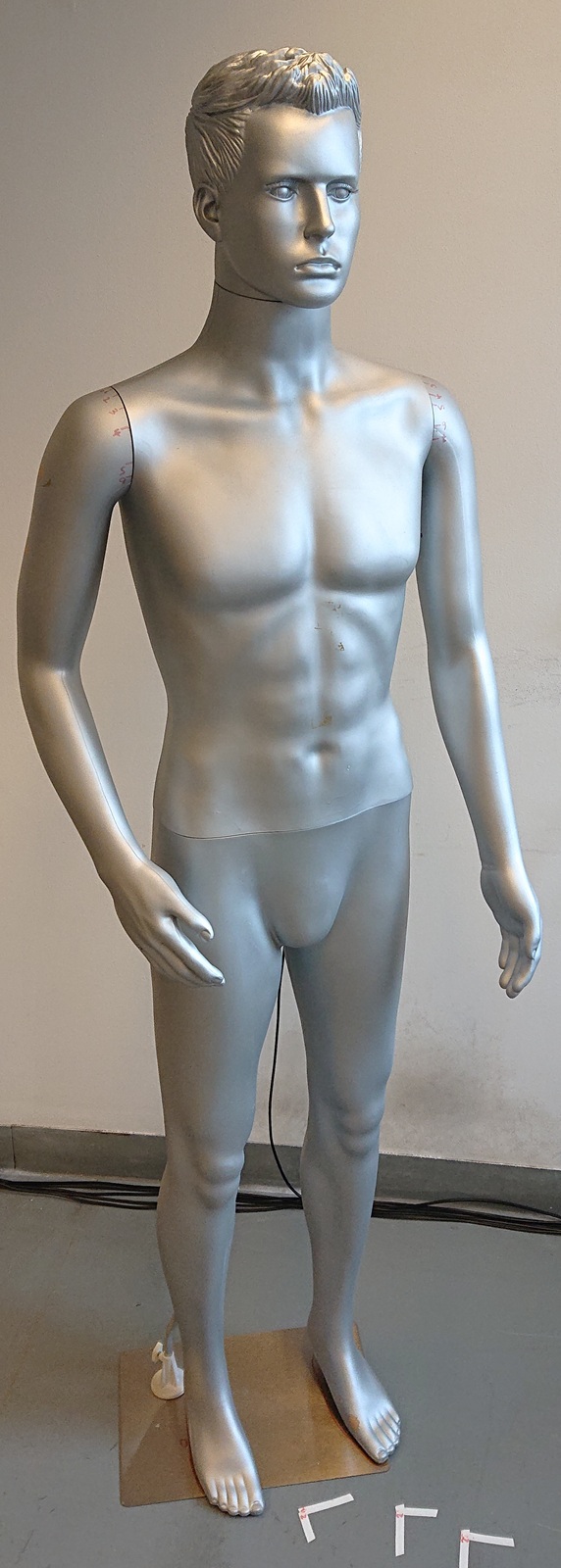}
		}
        \hspace*{-0.4cm}
        \subfloat[]{
            \includegraphics{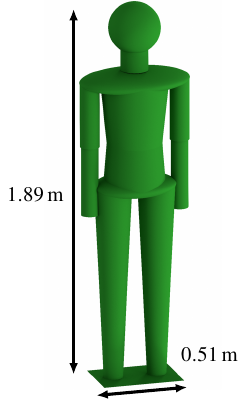}
		}

\caption{(a) 
Measurement setup with single Tx (left) and a receive patch antenna on a two-axis positioner (right). (b) Front view of the mannequin used in the acquisitions. (c) Simplified model of the phantom employed in simulations.}
\label{fig:measurement_setup}
\end{figure}

The test measurement setup~\cite{paulus_apwc25} is shown in Fig.~\ref{fig:measurement_setup}, and schematized in Fig.~\ref{fig:setup-schematic}. The transmitter (Tx) is placed \mbox{$0.10$\,m} from the wall behind at a height of \mbox{$1.0$\,m}, approximately centered along the wall. Reception is performed over a planar sampling grid mounted on a vertical frame, parallel to the opposite side wall and offset by \mbox{$0.283$\,m} from it. A single narrowband patch antenna is moved by a two-axis positioner to the grid locations, forming an effective receiving array with $162 \times 80$ samples (step \mbox{$3.1$\,cm}) over an area of about \mbox{$5$\,m} $\times$ \mbox{$2.5$\,m}. The room size is approximately \mbox{$5.6$\,m} $\times$ \mbox{$2.8$\,m} $\times$ \mbox{$3.6$\,m}; the nominal Tx–grid distance is about \mbox{$3.217$\,m}. Narrowband measurements are acquired around \mbox{$2.48$\,GHz} using a phase-stable link that provides the complex transmission $S_{21}$ between Tx and Rx (NanoVNA 6000-A~\cite{nanovna6000}).

\begin{figure}
  \centering
  \includegraphics[width=1\columnwidth]{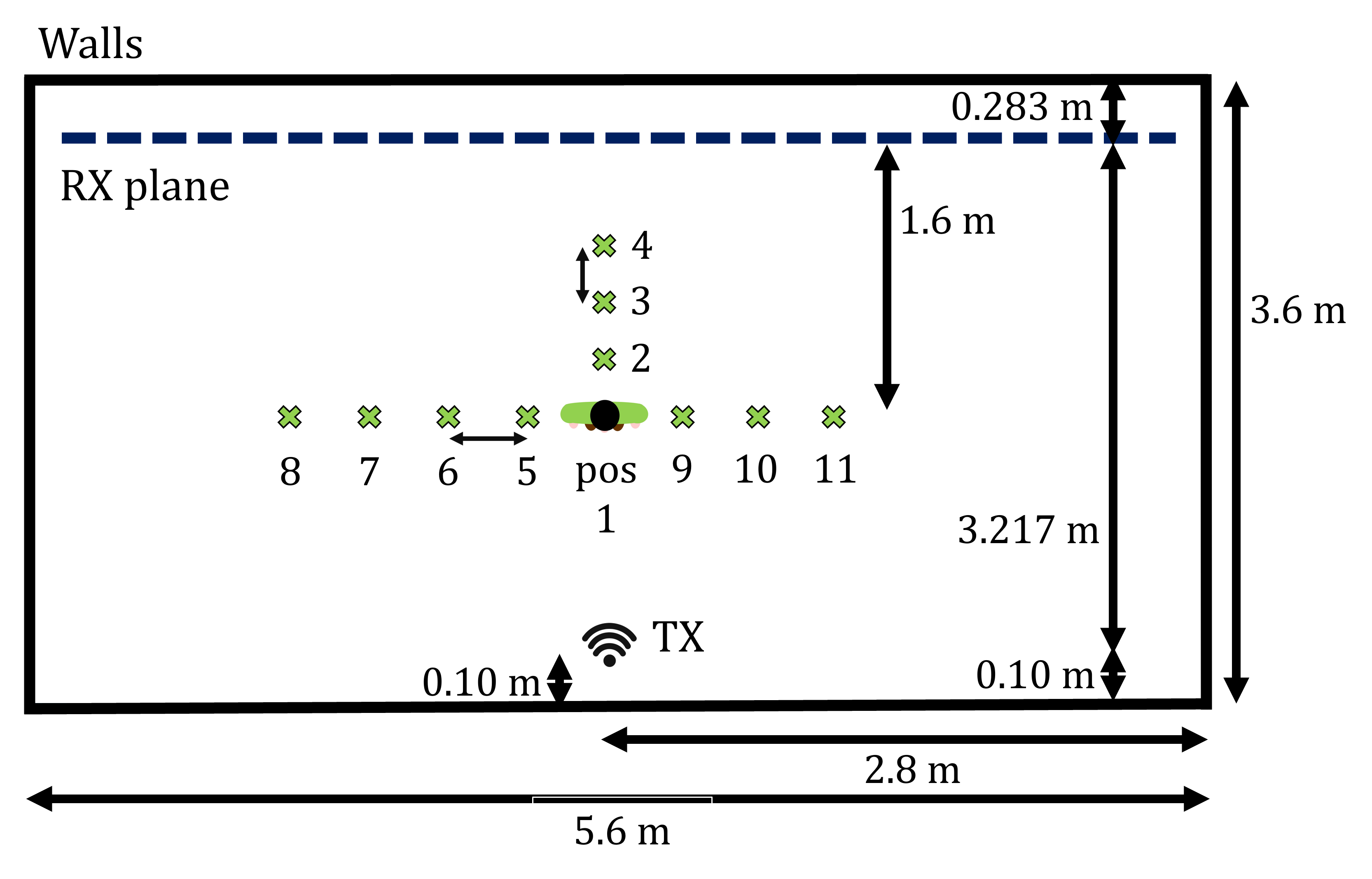}
  \caption{Plan view of the measurement environment with the transmitter Tx, the Rx sampling plane, and the representative target positions.}
  \label{fig:room-above}
\end{figure}

For device-free evaluation, the datasets include: (i) an \emph{empty-room} acquisition (no target), and (ii) acquisitions with a standing human phantom coated with zinc–aluminum paint and mounted on a \mbox{$40$\,cm} $\times$ \mbox{$29$\,cm} metal base. The mannequin, measuring \mbox{$1.89$\,m} in height and \mbox{$51$\,cm} in width, is used to emulate human presence. In representative trials it is placed at 10 positions within the room, as described in Fig.~\ref{fig:room-above}. Each grid point is sampled multiple times and averaged to reduce fast fading and instrument drift. Long scan times (about \mbox{$17$\,h} per dataset) are compensated by a simple drift correction on magnitude and phase (e.g., \mbox{$0.23$\,dB} and $7.8^\circ$) before computing differences.

Let $\overline{S}_{21}^{\mathrm{fp}}(m)$ and $\overline{S}_{21}^{\mathrm{tar}}(m)$ denote the averaged complex transmissions at receiver position $m$ in the empty-room and target-present cases. A magnitude-based attenuation map on the receiver grid is then defined as
\begin{equation}
M^{\mathrm{exp}}(m) = 20 \log_{10}\!\left(\frac{|\overline{S}_{21}^{\mathrm{fp}}(m)|}{|\overline{S}_{21}^{\mathrm{tar}}(m)|}\right)\;[\mathrm{dB}],
\label{eq:matt_meas}
\end{equation}
so that each pixel corresponds to one physical Rx position. This construction is consistent with radio-map practice in device-free systems and fingerprinting/tomography pipelines \cite{paper14,paper27,paper19}.

\begin{table}[t]
\caption{Measurement parameters}
\label{tab:meas_paras}
\centering
\begin{tabular}{c|c}
Room size & $5.6$\,m $\times$ $2.8$\,m $\times$ $3.6$\,m \\
Frequency & $2.4$–$2.5$\,GHz (slice at $2.48$\,GHz) \\
Scan plane & $\approx\,5$\,m $\times$ $2.5$\,m (vertical) \\
Step size & $\approx\,3.1$\,cm \\
Rx sampling & $162 \times 80 = 12960$ \\
Rx plane offset & $0.283$\,m from wall \\
Tx–grid distance & $\approx\,3.217$\,m \\
Scan time & $\approx\,17$\,h per dataset \\
\end{tabular}
\end{table}

\subsection{Full-Wave Simulation}
The simulation model was developed in the FEKO~\cite{Feko} simulation framework using the method of moments. A standard mesh size of $\lambda/12$ was adopted (with $\lambda$ the wavelength). Simulations were run at \mbox{$2.48$\,GHz} and considered the same target positions used in the measurement set. The transmitter was modeled as a single, center-fed electric dipole oriented vertically, and the vertical field component was sampled on the same Rx grid (patch pattern not explicitly modeled).

\section{Free-Space Validation}
\label{sec:free_space}

The first comparison assesses how well the full-wave simulation reproduces the measured field distribution in the simplest configuration, i.e., with no reflective boundaries and no target in the scene. In this free-space condition, only the transmitting dipole and the sampling grid are modeled, while walls, floor, ceiling, and mannequin are removed. The corresponding experimental reference is the empty-room acquisition, where the complex transmission coefficient $S_{21}(x,y)$ was recorded over the same grid without any obstruction between transmitter and receiver.

Since the measured $S_{21}$ represents the ratio of received to transmitted voltage at the instrument ports, it is proportional to the received electric field amplitude in the adopted setup. Therefore, the simulated field magnitude $|E(x,y)|$ and the measured $|S_{21}(x,y)|$ can be directly compared after normalization to their respective maxima to remove scale effects. 

\begin{figure*}[!t]
  \centering
    \includegraphics[width=0.85\textwidth]{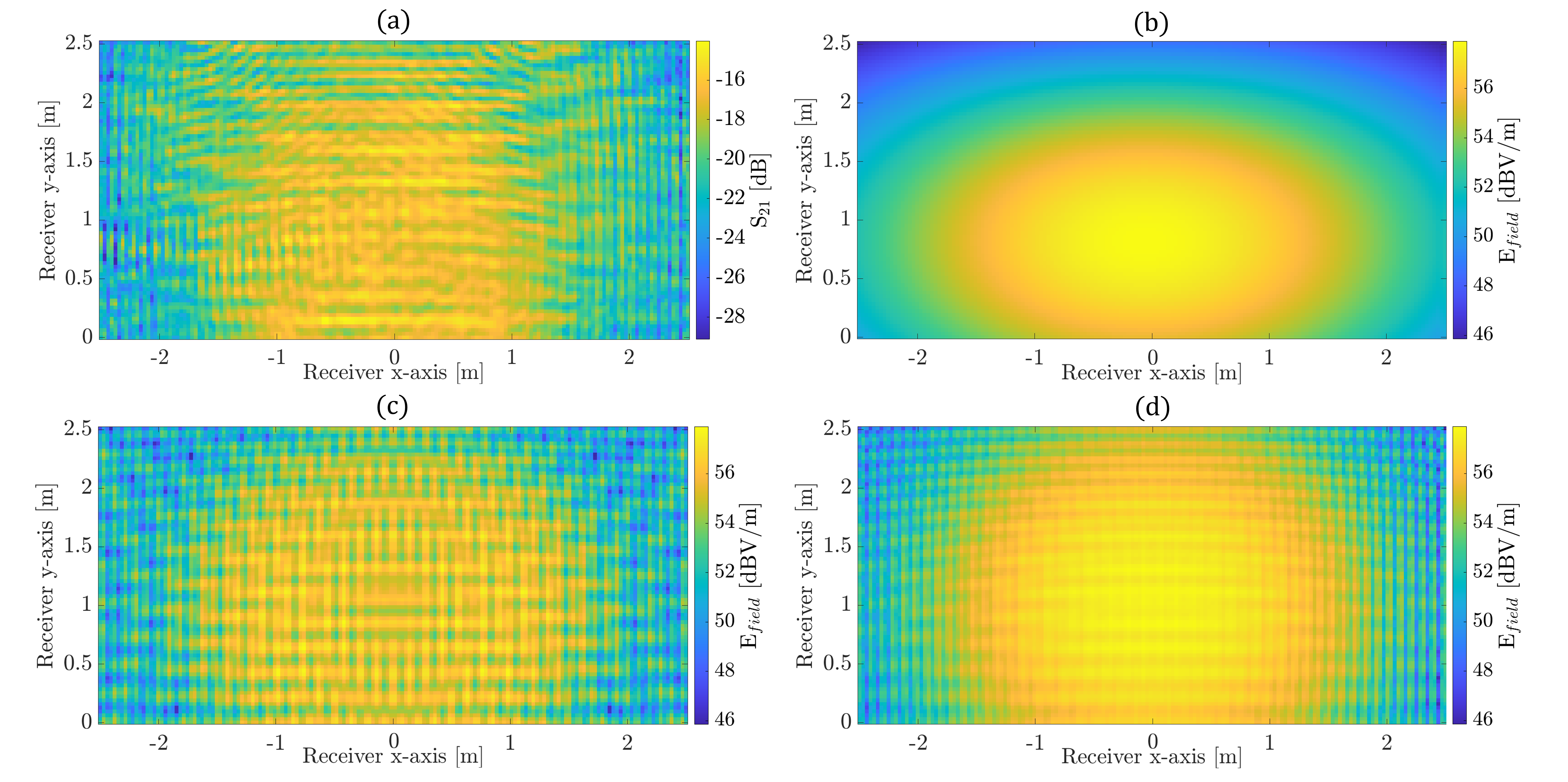}
  \caption{Free-space, frequency-averaged maps on the Rx grid. 
  (a) Measured $|S_{21}|$ in dB. 
  (b) Simulated $|E_\text{z,tot}|$ without walls (direct path only). 
  (c) Simulated $|E_\text{z,tot}|$ with wall/ceiling image sources using the initial reflection setup ($\Gamma$ as in Sec.~\ref{sec:method}). 
  (d) Simulated $|E_\text{z,tot}|$ with image sources and optimized reflection coefficients $\Gamma_i$ from the optimization.
  All maps are normalized to their own maxima before plotting to remove scale effects.}
  
  \label{fig:free-avg}
\end{figure*}

In what follows, map similarity is quantified using the Pearson correlation coefficient
\begin{equation}
\rho(A,B) = \frac{\sum_{i,j} (A_{ij} - \bar{A})(B_{ij} - \bar{B})}
{\sqrt{\sum_{i,j}(A_{ij} - \bar{A})^2}\sqrt{\sum_{i,j}(B_{ij} - \bar{B})^2}},
\label{eq:pearson}
\end{equation}
where $(i,j)$ index the grid sampling points, $A$ and $B$ are the simulated and measured field matrices, and $\bar{A}$, $\bar{B}$ are their spatial averages.

Because small geometric deviations (e.g., sub-wavelength Tx--Rx misalignments) can translate the interference pattern and artificially reduce $\rho$, a two-dimensional cross-correlation is computed and its maximum value over all possible shifts is retained as
\begin{equation}
\rho_{\max}(A,B) = \max_{\Delta x, \Delta y}\,
\rho\!\big(A,\mathrm{shift}(B;\Delta x,\Delta y)\big),
\label{eq:rhomax}
\end{equation}
where $\mathrm{shift}(B;\Delta x,\Delta y)$ denotes the matrix $B$ translated by $\Delta x$ and $\Delta y$ pixels.

Under the free-space conditions, the resulting correlation between the simulated and measured maps is $\rho_{\max}=0.55$. As observed in Fig.~\ref{fig:free-avg}(a)-(b), the direct line-of-sight propagation and the overall field decay are correctly reproduced by FEKO, but the fine-grained interference patterns observed in the measurements cannot be reproduced without including reflections from walls, floor, and ceiling.

\section{Reflection Modeling and Data-Driven Optimization }
\label{sec:reflections}

Multipath effects are now modeled by introducing wall and ceiling reflections using the image-source formulation described in~\cite[pp.~94--99]{Jin.2015}. Each wall is thus modeled by an image source scaled by a complex reflection coefficient $\Gamma = 0.203\,\mathrm{e}^{-\mathrm{j}13.5^\circ}$,
consistent with the dielectric properties of concrete~\cite{Fieramosca2023FloorEffects}. The inclusion of image sources leads to visible constructive and destructive fringes and a qualitative improvement in spatial agreement with measurements. While Fig.~\ref{fig:free-avg}(a)–(c) may suggest a comparable field distribution, this apparent similarity is largely influenced by the distinct color scales adopted. From a quantitative standpoint, however, the correlation increases only slightly, from $0.55$ to $0.57$. This suggests that the simplified assumption of a single, constant reflection coefficient does not fully capture the real indoor response.

To further refine the model, a data-driven optimization of the wall reflection coefficients was performed by exploiting the measured maps as reference. Because of the high computational cost of full-wave simulations, an auxiliary model was developed based solely on the image-source method, which can reproduce the dominant reflections with negligible runtime. This surrogate allows the use of numerical optimization routines to adjust the complex reflection coefficients $\Gamma_i$ (magnitude and phase) of each wall so as to maximize the correlation between simulated and measured fields.

The proposed model follows the simulation geometry: the transmitter is modeled as a Hertzian dipole with moment $p_0 = I_0 \ell = 1\,\mathrm{A{\cdot}m}$, while each wall is represented by a single image source. Multiple reflections are neglected due to their limited contribution ($|\Gamma|^n$ rapidly decays for $|\Gamma|<0.4$). The vertical field component $E_{z,\mathrm{tot}}(x,y,z)$ is computed at each receiver location as the coherent sum of the direct and reflected components as
\begin{equation}
E_{z,\mathrm{tot}}(x,y,z) = 
E_{z,0}(x,y,z) + \sum_{i=1}^{6} \Gamma_i E_{z,i}(x,y,z),
\end{equation}
where $E_{z,0}(x,y,z)$ is the direct contribution and $E_{z,i}(x,y,z)$ is  the contribution of the $i$-th image (i.e., $4$ walls, floor and ceiling). The optimization problem is formulated as
\begin{equation}
\label{problem}
\min_{\Gamma_i} \big[-\rho(S_{21},E_\text{z,tot})\big],
\end{equation}
where $\rho(\cdot,\cdot)$ is the Pearson correlation 
between the measured $S_{21}$ and the modeled $E_\text{z,tot}$ maps. Problem (\ref{problem}) is solved by using nonlinear programming implementation, with constraints $0 \leq |\Gamma_i| \leq 1$ and $0^\circ \leq \angle\Gamma_i \leq 360^\circ$, initialized at $|\Gamma|=0.2$.

The optimization yields the following effective coefficients:
\begin{align*}
\Gamma_{1} &= 0.19\angle 95^\circ (\text{right wall}), \quad
\Gamma_{2} = 0.15\angle 55^\circ (\text{left wall}),\\
\Gamma_{3} &= 0.11\angle 0^\circ (\text{ceiling}), \quad
\Gamma_{4} = 0.17\angle 17^\circ (\text{ground}),\\
\Gamma_{5} &= 0.11\angle 243^\circ (\text{back-Rx}), \quad
\Gamma_{6} = 0.70\angle 287^\circ (\text{back-Tx}).
\end{align*}
When applied to simulation, these optimized coefficients increase the correlation with the measured maps from $0.57$ to $0.67$. Averaging the measured data over the acquired frequency range to suppress noise yields even higher agreement: the correlation rises from $0.73$ (before optimization) to $0.82$ after optimization. These values confirm that the data-driven calibration of $\Gamma_i$ enables a realistic reproduction of indoor multipath patterns.

Overall, the proposed optimization approach, when combined with simulations, provides a computationally efficient method for parameter tuning. 
The agreement with experiments validate the use of this compact model as a reliable baseline for passive localization studies.

\section{Results and Discussion}
\label{sec:results}

\begin{figure*}[!t]
\centering
\includegraphics[width=\textwidth]{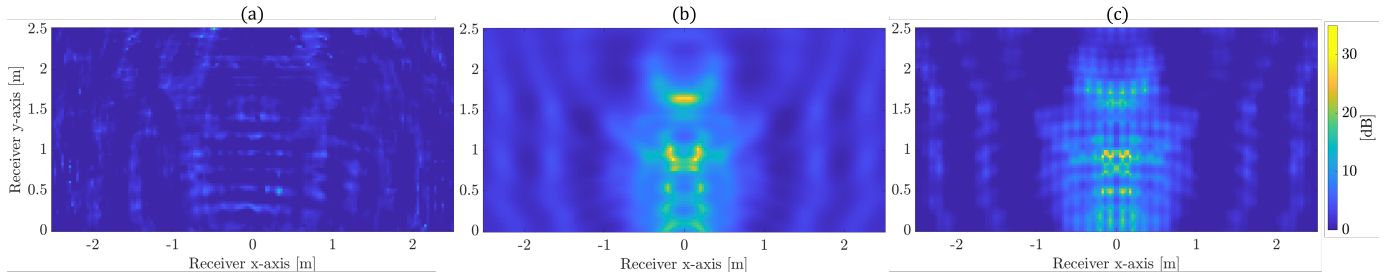}
\caption{Qualitative validation on the Rx grid with the target placed in line of sight (LoS) (position~1): (a) measured attenuation map (Eq.~\ref{eq:matt_meas}); (b) simulated map with floor only (no walls); (c) simulated map refined with image sources using the reflection coefficients $\Gamma_i$ calibrated in Secs.~\ref{sec:free_space}--\ref{sec:reflections}. 
Maps represent relative attenuation on the Rx grid and are compared in terms of spatial features rather than absolute levels. The refined map reproduces the main shadowing lobe, the fringe arrangement near the boundaries, and the standing-wave structure observed in measurements.
}
\label{fig:results}
\end{figure*}

Figure~\ref{fig:results} compares three attenuation maps (Eq.~\ref{eq:matt_meas}) for the target in position 1, on the line of sight (LoS) midway between Tx and the center of the Rx grid (Fig.~\ref{fig:room-above}): (a) the measured reference on the Rx grid, (b) the compact simulation with only the ground plane (no walls), and (c) the refined simulation that includes wall and ceiling interactions via image sources. In (c), the complex reflection coefficients \(\Gamma_i\) are those obtained in the previous sections from the free-space calibration (no target) and are reused \emph{as is}, i.e., without any re–tuning in the target-present case.

The floor-only model in panel~(b) reproduces the large-scale behavior but lacks the multipath structure clearly visible in the measurement. By injecting the calibrated image sources, panel~(c) recovers the salient qualitative features: (i) the width and location of the main shadow cast by the target along the LoS; (ii) the arrangement of constructive/destructive fringes in the vicinity of the side boundaries and towards the ceiling; and (iii) the vertical standing-wave pattern across the scanned aperture. The visual agreement holds not only in terms of lobe/null placement, but also in the relative spacing of the interference fringes, which matches the expected Fresnel geometry given the Tx–Rx separation and the target depth.

Residual differences remain but are consistent with the modeling simplifications and with the fact that we primarily target spatial feature consistency rather than absolute attenuation. The refined simulation slightly biases the absolute attenuation close to the mannequin silhouette, which is likely due to both the PEC approximation of the conductive coating and the use of idealized antenna patterns. Local discrepancies near recesses and around the target base suggest unmodeled edge diffraction and weak interactions with furniture/windows. Moreover, a single coefficient \(\Gamma_i\) per wall (independent of incidence and polarization) cannot fully capture angle-dependent behavior, nor higher-order reflections that 
may perturb fringe phasing near corners.

Despite these limitations, using the \(\Gamma_i\) calibrated in free space -without any retuning- proves sufficient to visually reproduce the key structures of the target-present maps. This confirms that a compact full-wave scene, augmented by image sources with data-driven coefficients, offers a practical and reusable approach for pre-measurement validation.

\section{Conclusion}
\label{sec:conc}

This work validated the capability of a compact, full-wave setup, implemented in FEKO, to reproduce realistic indoor propagation conditions and serve as a fast, pre-deployment reference for device-free localization studies. Starting from a minimal configuration—comprising a dipole transmitter, a sampled receiver grid, and no explicit walls or targets, the simulation was first validated against free-space measurements, showing a consistent large-scale field distribution but missing fine-grained interference patterns. The subsequent introduction of wall and ceiling interactions through image sources led to a qualitative improvement. Finally a data-driven optimization of the complex reflection coefficients~$\Gamma_i$ was performed using measurements of free space propagation (empty-room).

The optimized coefficients~$\Gamma_i$ were then reused, without further adjustment, to simulate the human-target case. 
Overall, the combination of full-wave simulation via FEKO and optimized reflection coefficients  provides a practical and computationally efficient framework for pre-deployment validation. The method reproduces realistic field distributions without a full CAD reconstruction, supporting the design of device-free sensing experiments.

Future work will be dedicated on extending the framework to multi-frequency validation and angle-dependent reflection coefficients, while integrating limited CAD for dominant features, realistic antenna patterns, and lightweight diffraction terms at critical edges.

\end{document}